\documentclass[12pt,preprint]{aastex}

\def\logz{\lbrack\hbox{M/H}\rbrack}

\slugcomment{To appear in the Astronomical Journal}
\shorttitle{Sextans A Star Formation History}
\shortauthors{Dolphin et al.}

\begin{document}

\title{Deep HST Imaging of Sextans A. III. The Star Formation History\linespread{1.0}\footnote{Based on observations with the NASA/ESA \textit{Hubble Space Telescope}, obtained at the Space Telescope Science Institute, which is operated by the Association of Universities for Research in Astronomy, Inc., under NASA contract NAS 5-26555. These observations are associated with proposal ID 7496.}}

\author{Andrew E. Dolphin, A. Saha}
\affil{Kitt Peak National Observatory, National Optical Astronomy Observatories, P.O. Box 26372, Tucson, AZ 85726}
\email{dolphin@noao.edu, saha@noao.edu}

\author{Evan D. Skillman, R.C. Dohm-Palmer}
\affil{Astronomy Department, University of Minnesota, Minneapolis, MN 55455}
\email{skillman@astro.umn.edu, rdpalmer@astro.umn.edu}

\author{Eline Tolstoy, A.A. Cole}
\affil{Kapteyn Institute, University of Groningen, P.O. Box 800, 9700AV Groningen, the Netherlands}
\email{etolstoy@astro.rug.nl, cole@astro.rug.nl}

\author{J.S. Gallagher, J.G. Hoessel}
\affil{University of Wisconsin, Dept. of Astronomy, 475 N. Charter St, Madison, WI 53706}
\email{jsg@astro.wisc.edu, hoessel@astro.wisc.edu}

\and

\author{Mario Mateo}
\affil{University of Michigan, Dept. of Astronomy, 821 Dennison Building, Ann Arbor, MI 48109-1090}
\email{mateo@astro.lsa.umich.edu}

\begin{abstract}
We present a measurement of the star formation history of Sextans A, based on WFPC2 photometry that is 50\% complete to $V = 27.5$ ($M_V \sim +1.9$) and $I = 27.0$.  The star formation history and chemical enrichment history have been measured through modeling of the CMD.  We find evidence for increased reddening in the youngest stellar populations and an intrinsic metallicity spread at all ages.  Sextans A has been actively forming stars at a high rate for $\sim 2.5$ Gyr ago, with an increased rate beginning $\sim 0.1$ Gyr ago.  We find a non-zero number of stars older than 2.5 Gyr, due to the limited depth of the photometry, a detailed star formation history at intermediate and older ages has considerable uncertainties.  The mean metallicity was found to be $\logz \approx -1.4$ over the measured history of the galaxy, with most of the enrichment happening at ages of at least 10 Gyr.  We also find that an rms metallicity spread of 0.15 dex at all ages allows the best fits to the observed CMD.  We revisit our determination of the recent star formation history (age $\le$ 0.7 Gyr) using BHeB stars and find good agreement for all but the last 25 Myr, a discrepancy resulting primarily from different distances used in the two analyses and the differential extinction in the youngest populations.  This indicates that star formation histories determined solely from BHeB stars should be confined to CMD regions where no contamination from reddened MS stars is present.
\end{abstract}

\keywords{galaxies: individual (Sextans A) --- galaxies: stellar content --- Local Group}

\section{Introduction}

Sextans A (DDO 75) is a dwarf irregular galaxy near the edge of the Local Group.  Its distance has recently been determined via a variety of methods to be $d = 1.32 \pm 0.04$ Mpc (Dolphin et al. 2003; hereafter paper II), which corresponds to a distance modulus of $\mu_0 = 25.61 \pm 0.07$ and places it in close proximity to Sextans B, NGC 3109, and the Antlia dwarf.

It has long been known that Sextans A is an active star-forming galaxy, with blue supergiants being observed in the early photometric work of \citet{san82} and HII regions observed by \citet{apa92} and \citet{hod94}.  Deeper CCD photometry by \citet{hoe83} and \citet{apa87} indicated that the current star formation rate was significantly higher than it had been in the past.

The first attempt to quantitatively determine the star formation history of Sextans A was done by \citet{doh97}, based on moderately deep WFPC2 observations.  They computed star formation histories based on the main sequence (MS) and blue helium-burning (BHeB) sequence luminosity functions.  They also used BHeB stars as age indicators (brighter means younger) to determine the spatially-resolved star formation history.  Similar work was done by \citet{van88}, who used wider-field but shallower data and found similar rates of recent star formation.  All of this work was limited to the past Gyr, leaving the question of the remainder of Sextans A's history open.

This is the third paper in a series based on new, deeper WFPC2 observations of Sextans A.  Paper I \citep{doh02} measured the recent star formation history (both global and spatially resolved) using the techniques of \citet{doh97}; the geometry of the two sets of WFPC2 data allow for such a study to be carried out over most of the galaxy.  In paper II, we studied the short-period variable star content and determined a new distance based on a variety of distance indicators.

In this paper, we seek to uncover the full star formation history of Sextans A through CMD modeling techniques (Dolphin 2002 and references therein).  Although such work would have been possible with the earlier WFPC2 data, the present data reach two magnitudes deeper, significantly below the red clump region.  In section 2, we describe the data and reductions.  Section 3 contains a ``manual'' determination of the star formation history by looking at the various components separately.  Finally, in section 4, we quantitatively measure the full star formation history using our CMD modeling techniques.

\section{Data and Reduction}

\subsection{Observations}

A detailed description of our observations is presented in paper I.  We summarize here.  As part of cycle 7 program GO-7496, we obtained 24 pairs of 1200s images with WFPC2.  Eight pairs of images were in F555W and sixteen pairs were in F814W.  The total integration times were thus 19200 seconds in F555W and 38400 seconds in F814W.  One quarter of the images were taken at each of four pointings, each offset by 0.25 arcsec, or approximately 5.5 PC pixels.  A 25th pair of images was taken in F656N, but will not be used for this paper.

\subsection{Photometry}

We retrieved the data from the STScI archive using on the fly calibration, which applies the best set of reference frames.  Reduction and photometry was made with the HSTphot package \citep{dol00a}.

To achieve the greatest depth and photometry accuracy, we cosmic ray cleaned all images in each filter and then co-added all images at the same pointing.  The result was eight deep images, corresponding to four pointings in each of two filters.  The procedure was similar to that used in paper II, except that we coadded images whenever possible here.  In paper II, we were looking for variable stars and thus kept images from different orbits separate even if the pointings were the same.

\begin{figure}
\plotone{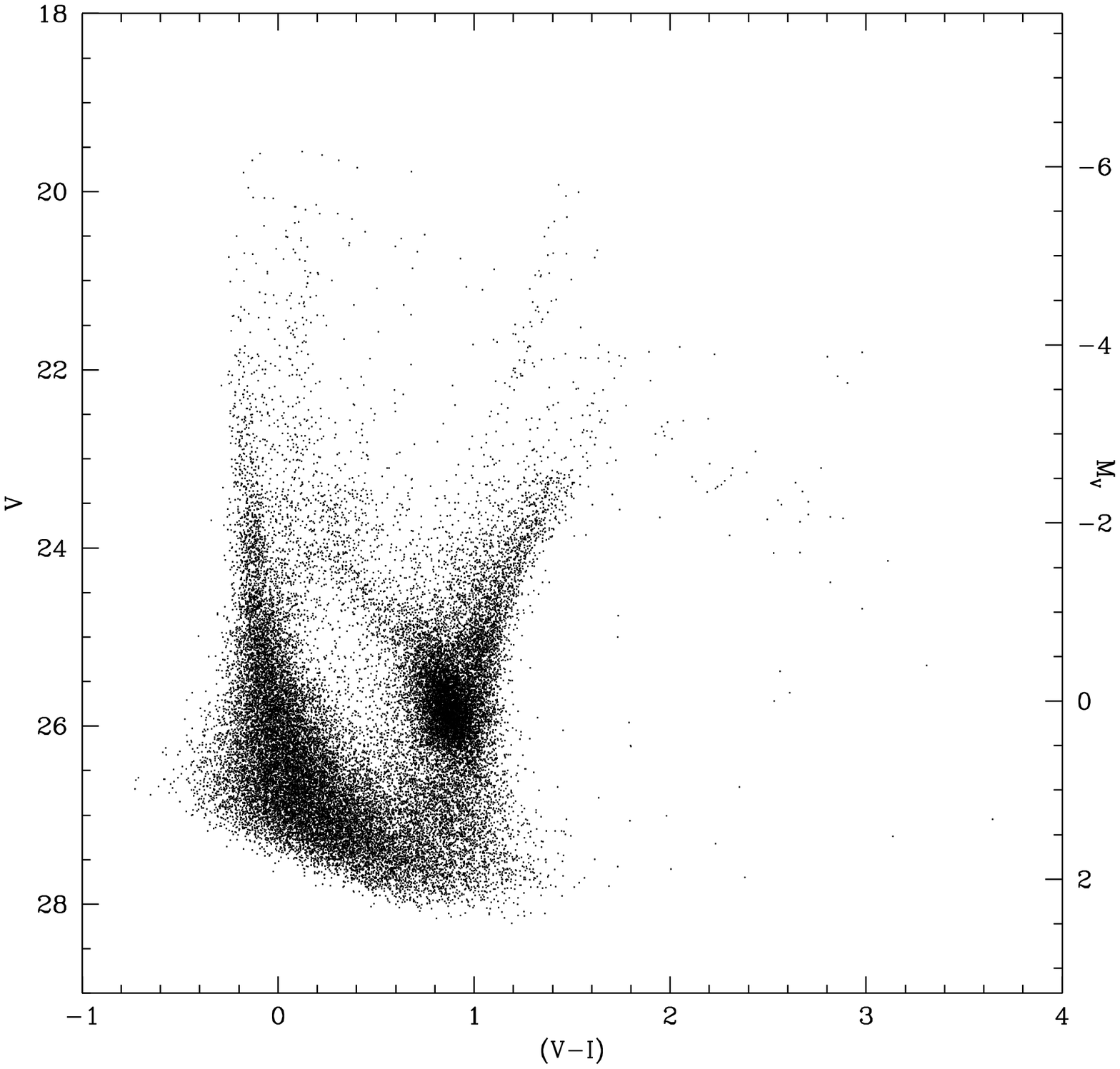}
\caption{$(V-I)$, $V$ Color-magnitude diagram (33340 stars). Poorly-fit stars ($\chi > 2.0$ or $|$sharpness$| > 0.3$) are not included.  Absolute magnitudes (on the right y-axis) are calculated assuming $V - M_V = 25.72$ \citet{dol03}. \label{figCMDc7}}
\end{figure}

HSTphot was used to simultaneously photometer stars on all eight deep images, producing combined $V$ and $I$ magnitudes in addition to magnitudes calculated from the separate images.  Calibration to the WFPC2 flight system was made using HSTphot-produced CTE corrections and calibrations \citep{dol00b}.  For this work, we consider only the combined magnitudes.  We measured a total of 42751 objects, of which 33340 passed our photometry cuts (observed in both filters, $\chi \le 2.0$, $|$sharpness$| \le 0.3$, and S/N$\ge 5$).  Figure \ref{figCMDc7} shows the CMD.
\placefigure{figCMDc7}

\begin{figure}
\plotone{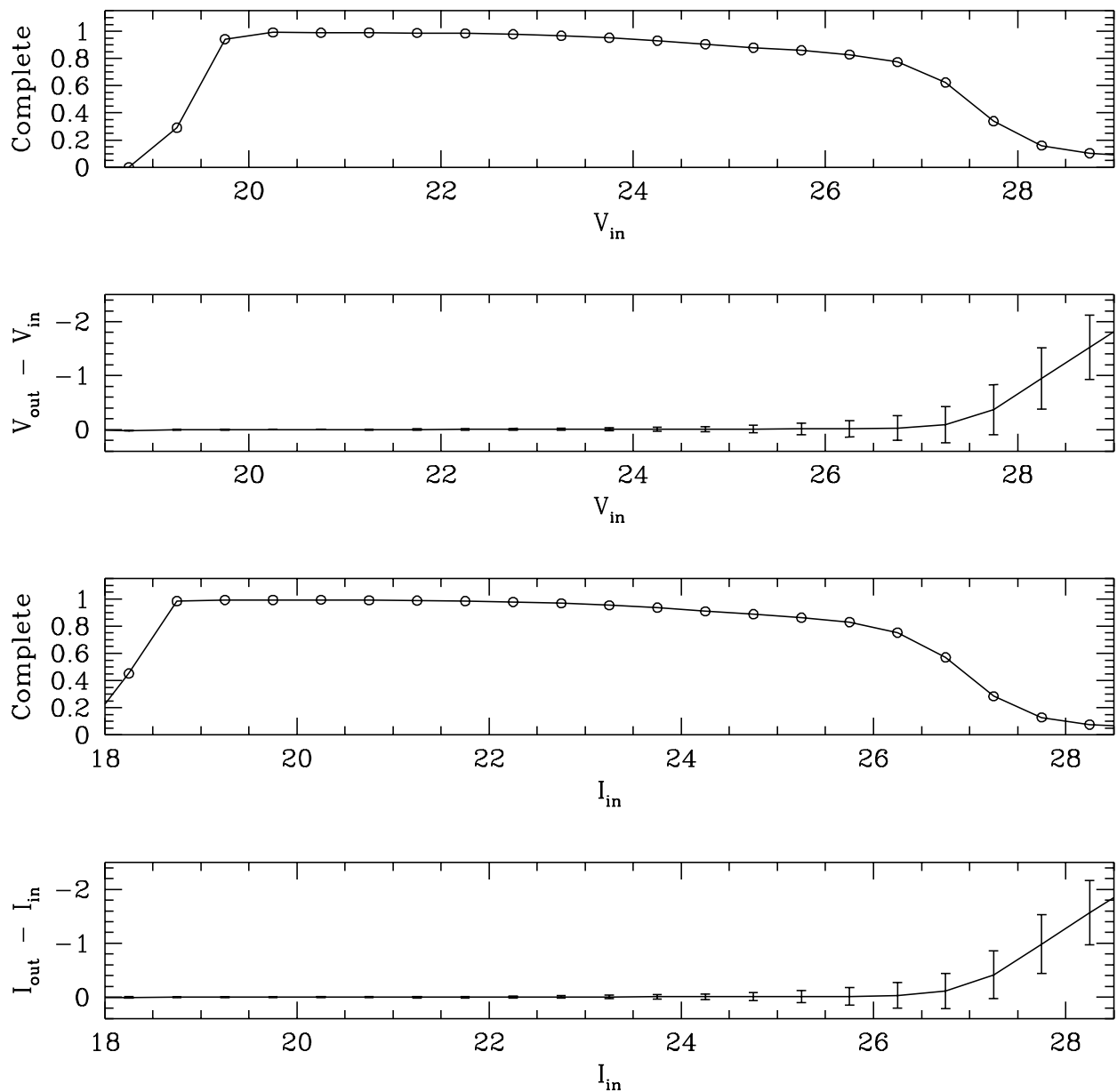}
\caption{Results of artificial star tests on our data.  The top panel shows the F555W ($V$) completeness as a function of input magnitude.  Note that the completeness never reaches 100\% because some objects are rejected for poor photometry at all magnitudes.  Incompleteness is poor at the bright end because of saturation, and at the faint end because of blending and signal-to-noise.  The second panel shows the mean F555W magnitude error (recovered minus input magnitudes) and rms scatter as a function of input magnitude.  Note that stars observed with magnitudes of $V > 27$ are likely to be much fainter stars.  The bottom two panels show the same data for F814W ($I$), for which our photometry is 0.5 magnitudes shallower. \label{figfakec7}}
\end{figure}

We made extensive artificial star tests over the color range $-0.5 < V-I < 3.0$ and the magnitude range $18 < V < 30$.  Stars were distributed on the images and on the CMD as functions of the observed star distribution, though care was made to ensure that all parts of the CMD adequately sampled.  In total, we made 1.36 million artificial star tests.  We show the results of the artificial star tests in Figure \ref{figfakec7}.
\placefigure{figfakec7}

\section{Stellar Populations}

Because the CMD of Sextans A contains stars of all ages, from recently-formed stars on the MS to ancient stars on the red giant branch (RGB), one can use the CMD to determine the star formation history of the galaxy.  In nearby systems, one can simply make an unconstrained maximum likelihood solution for all unknowns: distance, extinction, SFR(t), and Z(t) using the techniques described by Dolphin (2002 and references therein).  However, \citet{dol02} has demonstrated that one needs photometry extending well below the horizontal branch (i.e., at least to $M_V = +2.0$) in order to constrain the oldest stellar populations.  Unfortunately, this a luxury we do not have for galaxies at the distance of Sextans A.  Thus we need to first examine the components of the CMD in turn before we can determine a meaningful star formation history.  We will do so in order of importance on the CMD.

\subsection{Young Stars (age $<$ 1 Gyr)}

We first turn our attention to the young stars in Sextans A, which are seen in the CMD on the MS and BHeB sequence.  Paper I presented a detailed derivation of the recent star formation history using these stars, so we will discuss them only briefly here.

\begin{figure}
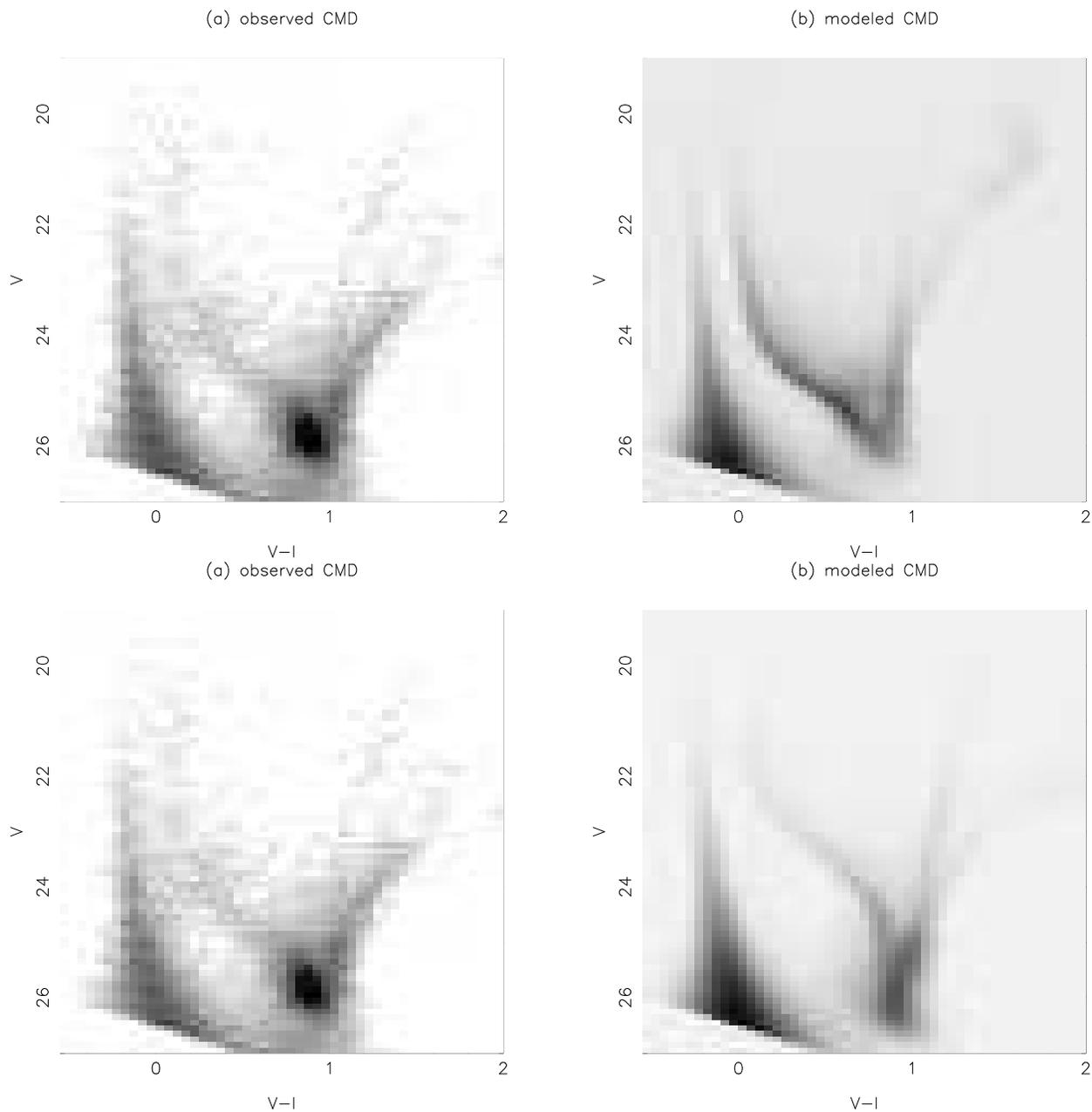

\plotone{dolphin.fig03a.ps}
\plotone{dolphin.fig03b.ps}
\caption{A comparison of the observed and synthetic CMDs in order to see the effect of changing metallicity, particularly with regard to the position of the BHeB sequence.  The top two panels show the observed CMD (left) and a synthetic CMD calculated using a metallicity of $\logz = -1.7$ (right).  The bottom panels are the same, but the synthetic CMD was calculated using a metallicity of $\logz = -0.7$.  In both cases, the synthetic CMDs were calculated using the Padova models \citep{gir00} and our artificial star tests. \label{figYCMD0}}
\end{figure}

Theoretical stellar evolution models predict that the position BHeB sequence is a strong function of metallicity.  Using the Padova models \citep{gir00}, we have created synthetic CMDs of young stars for metallicities Z=0.0004 ($\logz = -1.7$) and Z=0.004 ($\logz = -0.7$); these are shown and compared with the observed CMD in Figure \ref{figYCMD0}.  In preparing these figures, we adopted the distance ($\mu_0 = 25.61$) and extinction ($A_V = 0.12$) determined in paper II and a power law IMF with a slope of $\Gamma = -1.30$ (compared with a Salpeter value of $\Gamma = -1.35$).  Note that we intentionally do not plot isochrones over the observed CMD.  The MS and BHeB sequence both have finite widths due to evolutionary effects which must be accounted for in modeling the CMD.  Thus, simple line representations of these populations can be misleading.
\placefigure{figYCMD0}

Comparing the observed and synthetic CMDs, it is clear that the BHeB sequence in the $\logz = -0.7$ CMD falls well above the observed data, while that in the $\logz = -1.7$ sequence falls along the lower edge of the observed sequence.  We note that both synthetic CMDs have sharper BHeB sequences than the observed data; we interpret this as evidence of a metallicity spread.  Fitting the BHeB sequence position and spread, we find that the observed CMD is best fit by a current mean metallicity of $\logz = -1.45$, with a metallicity spread of $\pm 0.20$ dex.  Note that if the errors in our photometry were significantly underestimated, then the observations would be consistent with a smaller metallicity spread.

However, we also find that the synthetic MS is too narrow to fit the observed data, regardless of the adopted metallicity (or metallicity spread).  There are several possible explanations for this.  One is that the isochrones are grossly in error on the main sequence, a possibility whose detailed exploration is beyond the scope of our paper.  A second possibility would be the presence of a greater metallicity spread.  However we note that $V-I$ colors of upper main sequence stars change by only $\sim 0.03$ magnitudes per dex of metallicity; thus even a huge metallicity spread would not explain the main sequence color spread we observe.

A more plausible option is the presence of differential extinction in the youngest populations, caused by the preferential proximity of young stars to dusty star-forming regions.  Such an effect has been observed in the LMC by \citet{har97}, who find extinction values in excess of $A_V = 1.5$ for OB stars but very few red giants with extinction values in excess of $A_V = 0.6$, based on their $UBVI$ photometric survey.  As Sextans A is also actively forming stars, one should expect to see such an effect here as well, albeit with lower overall extinction values due to its lower metallicity.  This conclusion is strengthened by the photometry of \citet{apa87}, who found an excess number of red points in their $(U-B)$, $(B-V)$ diagram that they interpreted as evidence of internal reddening.

Determination of the exact functional form of the extinction distribution is a more difficult task.  We note that the Cepheid period-magnitude relations from paper II showed no sign of differential extinction, and thus assume no differential extinction at ages $\ge 500$ Myr.  We also note that the width of the red helium-burning sequence fainter than $V = 21.25$ ($M_V = -4.5$) is explained solely by the small metallicity spread noted above; this implies little differential extinction at ages $\ge 70$ Myr.  The MS shape is best fit with some differential extinction for ages younger than $\sim 90$ Myr and significantly higher amounts at ages younger than $\sim 40$ Myr.

To model the differential extinction as simply as possible (given the constraints above), we apply a flat distribution of differential extinction values to all stars younger than 100 Myr, with the maximum extinction increasing linearly from zero at an age of 100 Myr to 0.5 magnitudes in $A_V$ at an age of 40 Myr.

\begin{figure}
\plotone{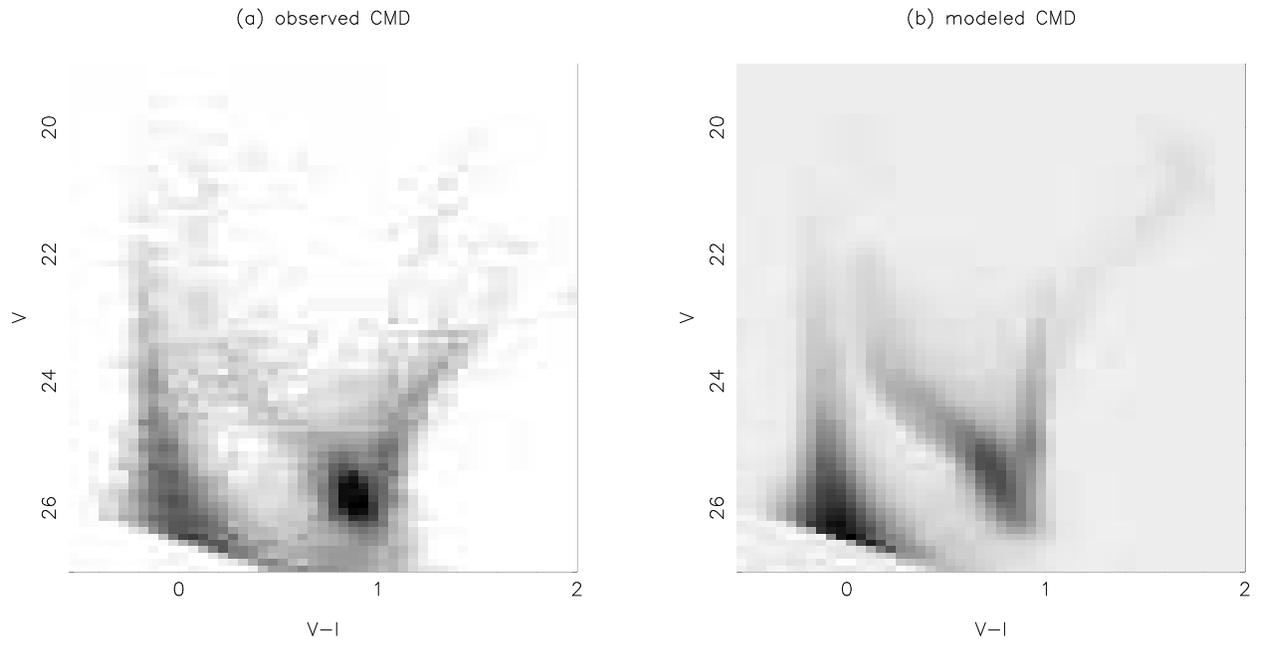}
\caption{Like Figure \ref{figYCMD0}, but using a range of metallicities from $-1.65 < \logz < -1.25$.  Differential extinction was added to better fit the color and color spread of the MS. \label{figYCMD}}
\end{figure}

Incorporating all of these features (current metallicity, metallicity spread, and differential extinction) into the CMD produces the ``young'' synthetic CMD shown in Figure \ref{figYCMD}.
\placefigure{figYCMD}

\subsection{Old Stars (age $>$ 2 Gyr)}

\begin{figure}
\plotone{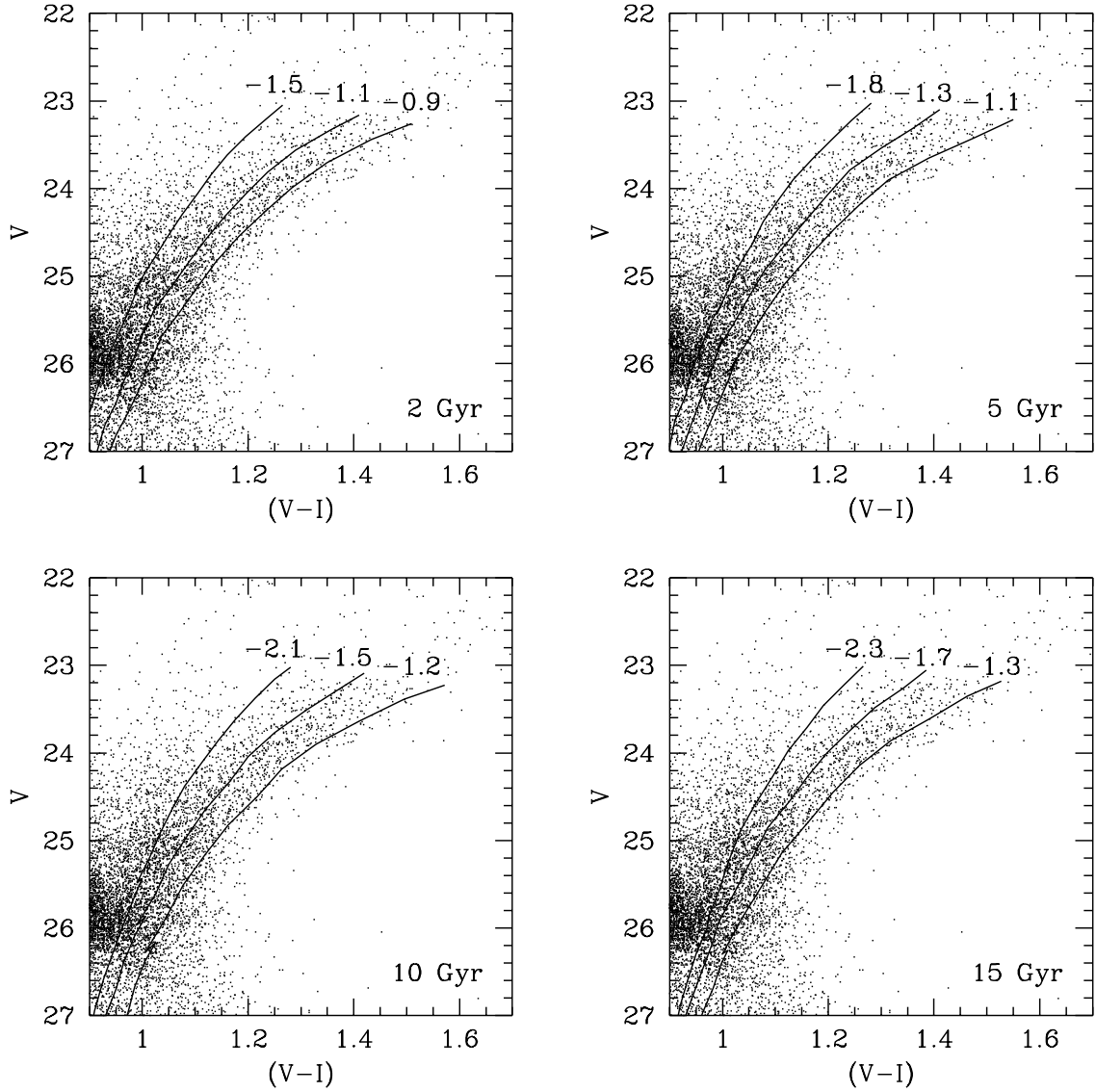}
\caption{Constraints on the age-metallicity relation, as determined from RGB stars.  The four panels show the RGB portion of the CMD, along with three isochrones for each age.  Note that the three isochrones are in roughly the same positions in each panel, but have different metallicities.  This is an example of the age-metallicity degeneracy. \label{figagez}}
\end{figure}

The dominant CMD feature that is not caused by young stars is the RGB; we now turn our attention to that.  Although there are some younger stars falling in that part of the CMD, the RGB is predominantly composed of stars older than $\sim 2$ Gyr.  The position of the RGB is a function of age and metallicity, in that a star that is older or more metal-rich will be redder.  We show the range of possible metallicity values for 2, 5, 10, and 15 Gyr stars in Figure \ref{figagez}, calculated from the \citet{gir00} theoretical isochrones.
\placefigure{figagez}

The blue edge of the RGB is poorly-defined, as there are significant numbers of population I and AGB stars in that part of the CMD.  However, the red edge is fairly sharp, and thus provides strong constraints for the metallicity of RGB stars.  The red edge can be fit with 2 Gyr old stars of metallicity $\logz \approx -0.9$, 15 Gyr old stars of metallicity $\logz \approx -1.3$, or stars with intermediate ages and metallicities.  In the previous section we determined a metallicity of $-1.45 \pm 0.20$ for the young stars.  Thus, we would expect very few old stars to have metallicity higher than this.  From Figure \ref{figagez}, we can conclude that the stars at the red edge of the RGB are likely old stars (ages $\ge$ 10 Gyr), with metallicity $\logz \approx -1.3$.

We note that the presence of old stars with metallicities similar to young stars implies that Sextans A reached its current metallicity very quickly.  Such a feature is not unique to Sextans A; other metal-poor galaxies with extended star formation such as Leo I \citep{dol02} and UGC 4438 \citep{dol01} have shown little metal enrichment over most of their histories.  This is not necessarily inconsistent with the bulk of their star formation occurring at later ages.  In the cases of both Sextans A and UGC 4438, their present day metallicities (12 $+$ log(O/H) $\approx$ 7.5, Skillman et al.\ 1989; 1994) are not that much higher than that seen almost universally at high redshift \citep{pet99}.  A reasonable picture of Sextans A is thus that it formed from pre-enriched gas, and that most of the metals produced by its stars have escaped.

\begin{figure}
\plotone{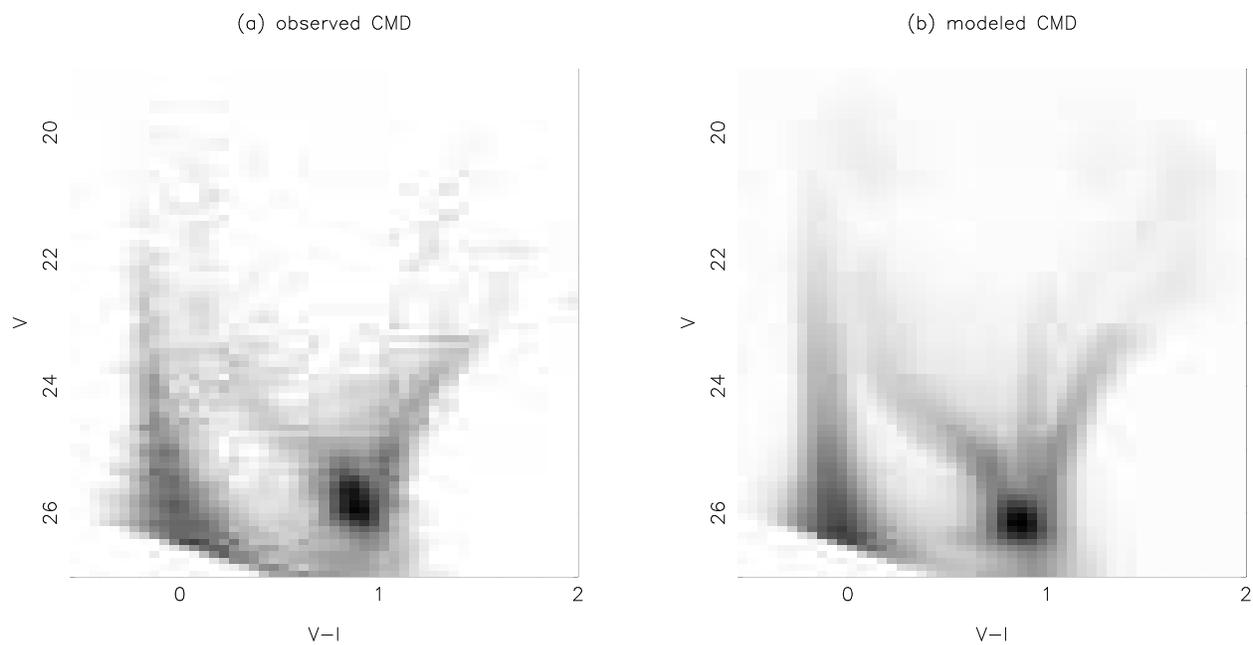}
\caption{A comparison of observed and synthetic CMDs.  The synthetic CMD contains both young ($<1$ Gyr) and old ($>2$ Gyr) stars, but not stars of intermediate ages.  Metallicities span the range $-1.75 < \logz < -1.35$ for all ages.  Panels are the same as in Figure \ref{figYCMD0}. \label{figYOCMD}}
\end{figure}

To account for the young and old stars, we will adopt a mean metallicity of $\logz = -1.45$ with spread of $\pm 0.20$ dex for all ages.  The synthetic CMD (still not containing stars between $1$ and $2$ Gyr) is shown in Figure \ref{figYOCMD}.
\placefigure{figYOCMD}

The observed CMD (Figure \ref{figCMDc7}) shows a feature that extends redwards from the RGB tip to $(V-I) \approx 3$.  Modeling this using the \citet{gir00} isochrones would require a metallicity of $\logz = -0.6$ at an age of 3 Gyr ago, which is inconsistent with our observations of the metallicity of younger stars and would require an RGB tip color of $V-I = 1.95$ (about half a magnitude redder than what is observed).  We conclude that these stars are a combination of TP-AGB and Carbon stars, evolutionary phases that are observed at this color and absolute magnitude in other metal-poor galaxies such as IC 1613 \citep{alb00}.

Finally, we note that there is no clearly-delineated horizontal branch in Sextans A, a result of the large number of young stars in that part of the CMD.  Consequently a measurement of the number of stars $> 10$ Gyr old can only be done via statistical modeling of the CMD.

\subsection{Intermediate-age Stars}

\begin{figure}
\plotone{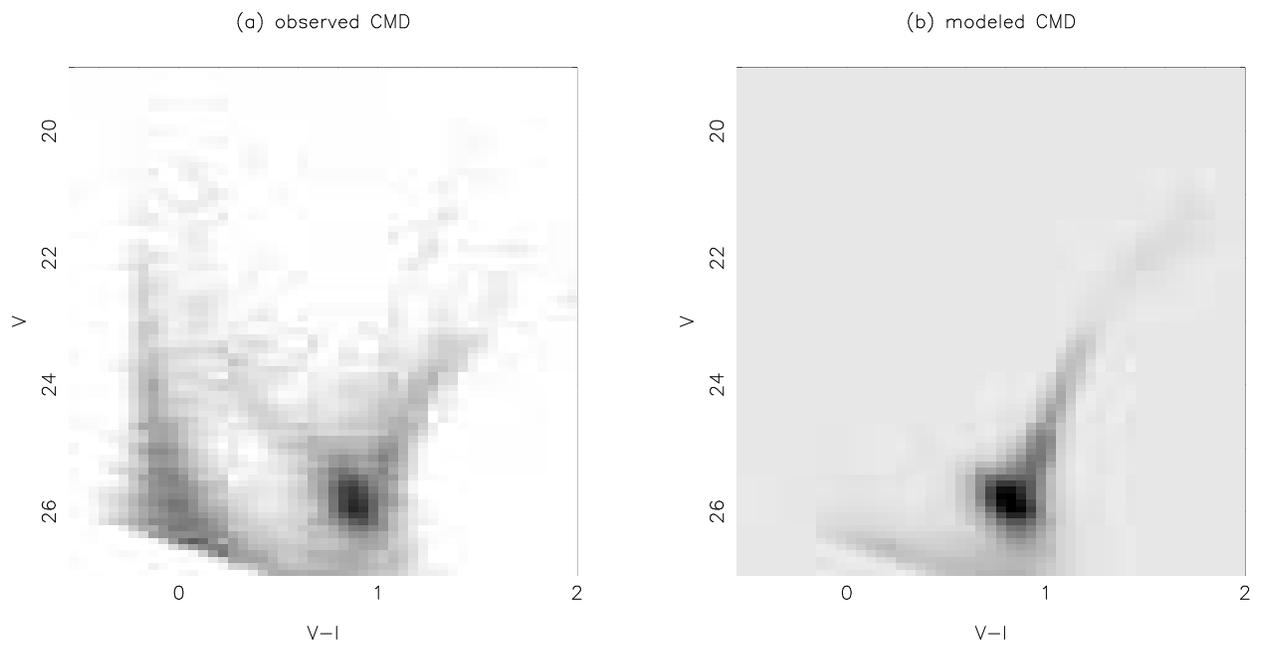}
\caption{A comparison of observed and synthetic CMDs.  The synthetic CMD contains only stars with ages between 1 and 2 Gyr.  Metallicities span the range $-1.75 < \logz < -1.35$ for all ages, as with the young and old stars shown in Figure \ref{figYOCMD}.  Panels are the same as in Figure \ref{figYCMD0}. \label{figINTCMD}}
\end{figure}

The final stars to be considered are the stars with ages between 1 and 2 Gyr.  Although ``intermediate-age'' has many meanings, we adopt the practical definition of being older than the oldest BHeB stars but younger than the stars on the RGB.  The synthetic CMD with just these stars is shown in Figure \ref{figINTCMD}.  It is clear from the figure that these stars contribute to the red clump, upper AGB, and lower MS; however disentangling them from the younger and older populations is not trivial because they do not occupy otherwise-empty parts of the CMD.  The statistical decomposition of the CMD will be the topic of the next section.
\placefigure{figINTCMD}

\section{Star Formation History}

\begin{figure}
\plotone{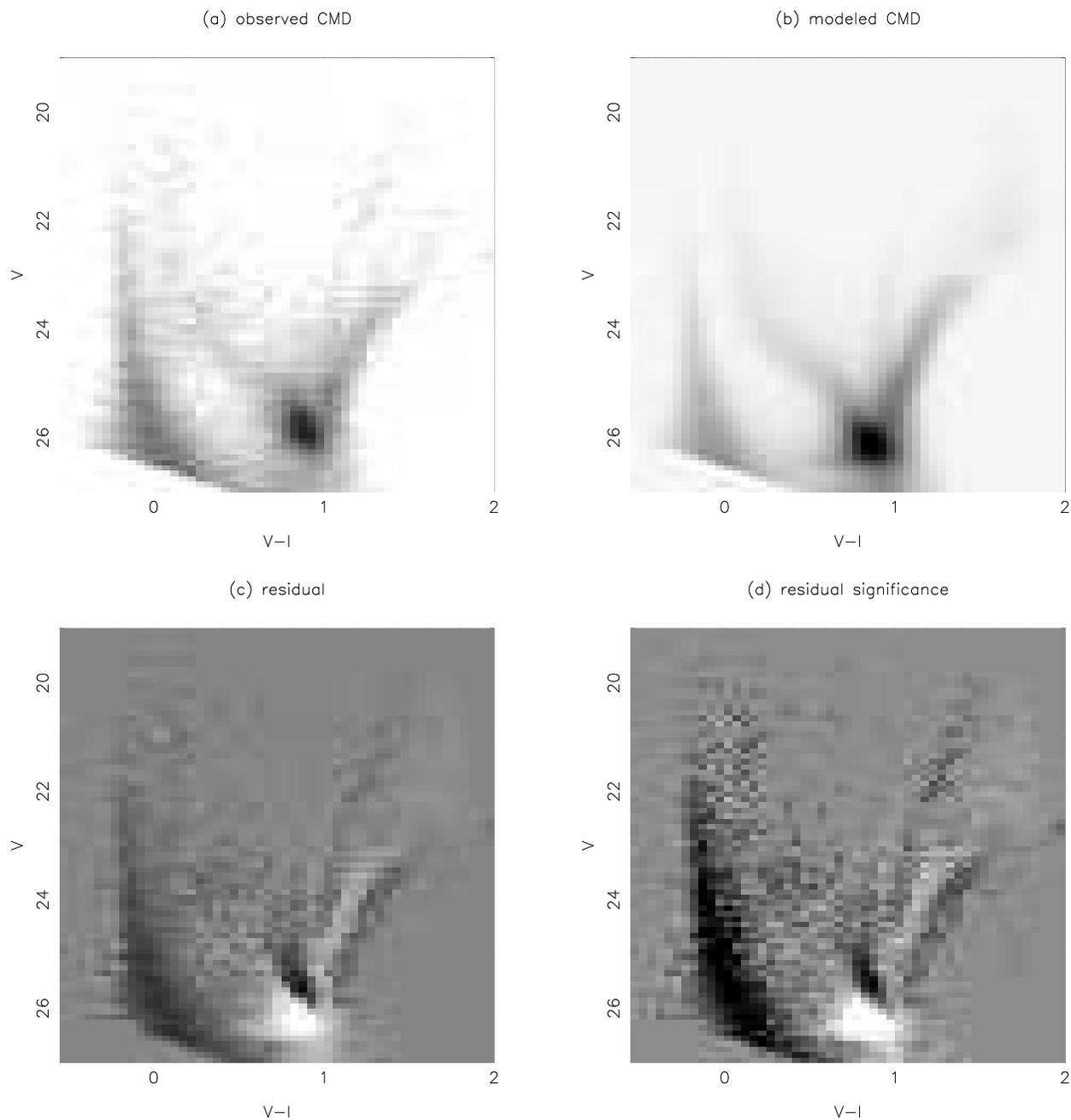}
\caption{A comparison of observed and synthetic CMDs assuming a constant star formation rate for the lifetime of the galaxy.  Metallicities span the range $-1.75 < \logz < -1.35$ for all ages.  The top panels are the same as in Figure \ref{figYCMD0}. The bottom panels show the residuals (left) and the significance of the residuals (right).  The black residual in the main sequence signifies many more observed stars than synthetic stars; the white residual in the lower red clump indicates an excess of synthetic stars.  This implies that the current star formation rate is much higher than the ancient rate, and thus the assumption of a constant star formation rate is untenable. \label{figCMDconst}}
\end{figure}

Armed with the components of the CMD that are produced by stars of a variety of ages, we set out to determine what combination of these components best fits the observed data.  The simplest hypothesis is that the star formation rate has been constant over the history of the galaxy.  A synthetic CMD computed with such a history is shown in Figure \ref{figCMDconst}, adopting the metallicity and extinction from the previous section.  From the difference between the observed and synthetic diagrams, it is clear that the synthetic diagram has too many old stars (from the oversubtraction of the RGB and lower red clump) and too few young stars (from the undersubtraction of the MS and BHeB sequence).  Put differently, the star formation rate over the past Gyr or two is significantly higher than that $> 5-10$ Gyr ago.
\placefigure{figCMDconst}

Because of the failure of the constant SFR model, a more sophisticated fit for the history is necessary.  The statistical basis behind this procedure and tests of its accuracy were given by \citet{dol02}; we merely summarize here.  We produce a large number of ``single-population'' synthetic CMDs, each containing only stars with limited ranges (0.1 dex) of age and metallicity.  These CMDs are generated using a combination of theoretical isochrones, IMF, binary function, distance, and extinction distribution.  Observational errors (from photon noise and blending) are simulated by using the completeness, photometric bias, and photometric scatter (all functions of color and magnitude) measured in the artificial star tests.

These synthetic CMDs, as well as simulated CMDs of foreground stars and bad points, are combined linearly to calculate the expected distribution of stars on the CMD for any star formation history.  With synthetic and observed CMDs in hand, we calculate the likelihood that the observed data were produced by the star formation history used to produce that particular synthetic CMD; repeating this procedure for many histories to maximize the likelihood allows us to measure the history of Sextans A.

As with the \citet{tol98} study of Leo A, we find that the CMD is better-fit with a shorter distance than with our preferred distance of $\mu_0 = 25.61$.  We attribute this to uncertainties in the theoretical models of high-mass stars, and make our solution using the longer distance.  To compensate for potential numerical instabilities caused by the isochrones, we ran three separate solutions -- one in the normal way described by \citet{dol02}, one at low resolution (to emphasize the numbers of stars in each part of the CMD while deemphasizing positions of features), and one excluding the red clump region from the CMD (to reduce the reliance of the solution on that dominant feature).  Each of the three solutions was consistent with the others, giving us confidence in the robustness of our results.

Sextans A appears to have had a significant event $1-2.5$ Gyr ago, when its star formation greatly increased.  Star formation has continued since that event, with another increase of a factor of three $0.06-0.1$ Gyr ago.  It is unclear what triggered the sudden increases in star formation.  These events are frequently linked with interactions with other galaxies, and Sextans A has a very close companion, Sextans B.  Interestingly, based on the HIPASS observation of a smooth HI distribution around Sextans A, Wilcots \& Hunter (2002) have argued that Sextans A has {\it not} interacted with another galaxy in the last 6 -- 10 Gyr.  Since the global dynamics of the HI associated with the optical galaxy show large distortions (i.e., misalignment of optical and kinematic axes, Skillman et al. 1988), and the HIPASS observations are very low angular resolution, perhaps a recent interaction with Sextans B is not ruled out, and possibly even likely.

The star formation history at older ages is less certain.  We believe the star formation rate to have been much lower than at present, but nonzero.  However, because the old stars are dominated by the large numbers of younger stars, it is impossible to quantify the star formation rate or give detailed properties of the old population.

Confirming what we found in the previous section, we measured a roughly constant (to within 0.1 dex) metallicity of $\logz = -1.4$ over the measurable history of the galaxy.  The CMD was best fit with an rms spread of 0.15 dex, a spread present at all ages.  We note that the recovered metallicity is consistent with that of $\sim 4\%$ solar determined by \citet{ski89} from HII region spectroscopy.

\begin{figure}
\plotone{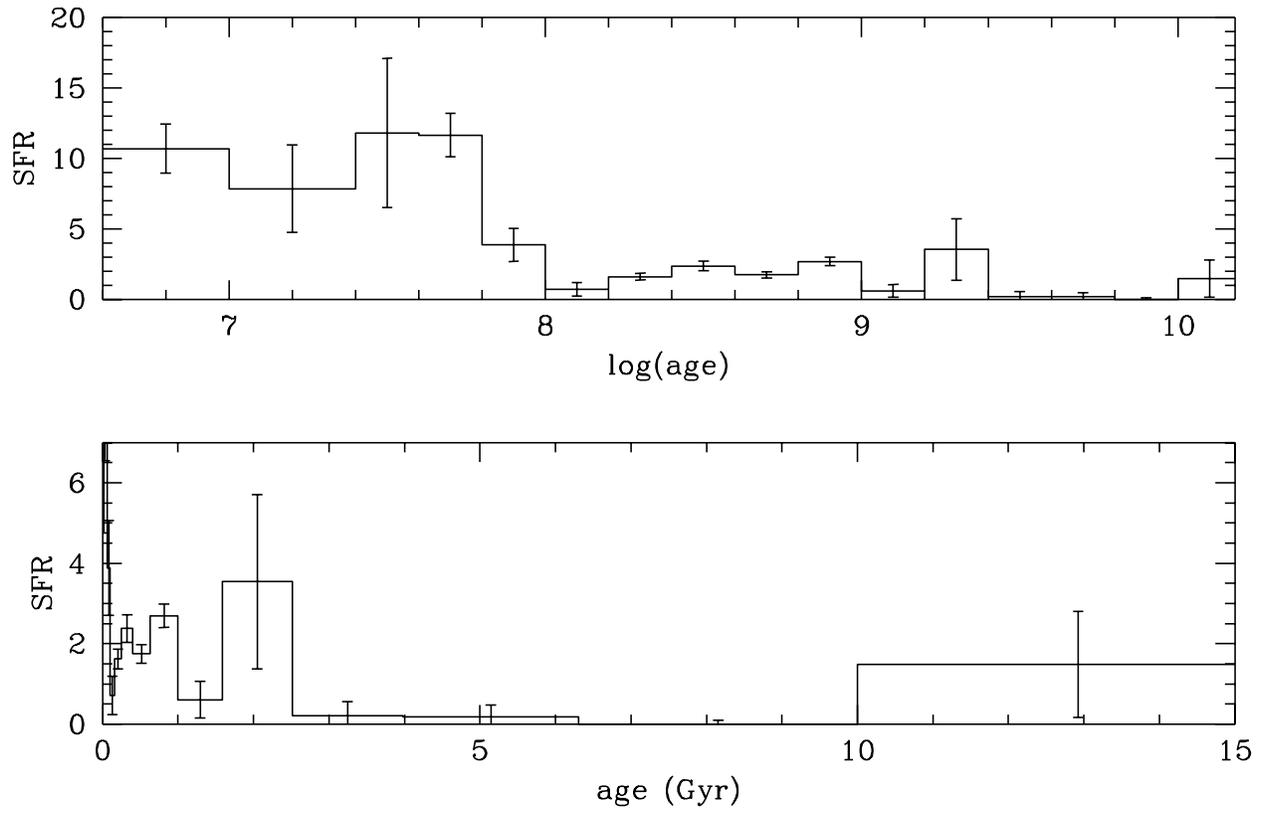}
\caption{Star formation history of Sextans A.  The top panel is on a logarithmic time scale; the bottom panel has a linear scale.  Both panels show the star formation history normalized to a lifetime average of 1.0. \label{figSFH}}
\end{figure}

We show the star formation history in Table \ref{tabSFH} and Figure \ref{figSFH}, normalized to a lifetime average of 1.  A Monte Carlo test was used to determine the statistical uncertainties; systematic errors were estimated by comparing our three solutions.  Conversion of our relative star formation rates into absolute rates depends heavily on the assumed IMF for low-mass stars.  Assuming the IMF of \citet{kro01}, we calculate a lifetime average star formation rate of $600 \pm 300 M_{\odot} Myr^{-1} kpc^{-2}$ for this field, and an average rate of $1600 \pm 500 M_{\odot} Myr^{-1} kpc^{-2}$ during the past 2.5 Gyr.
\placetable{tabSFH}
\placefigure{figSFH}

\begin{figure}
\plotone{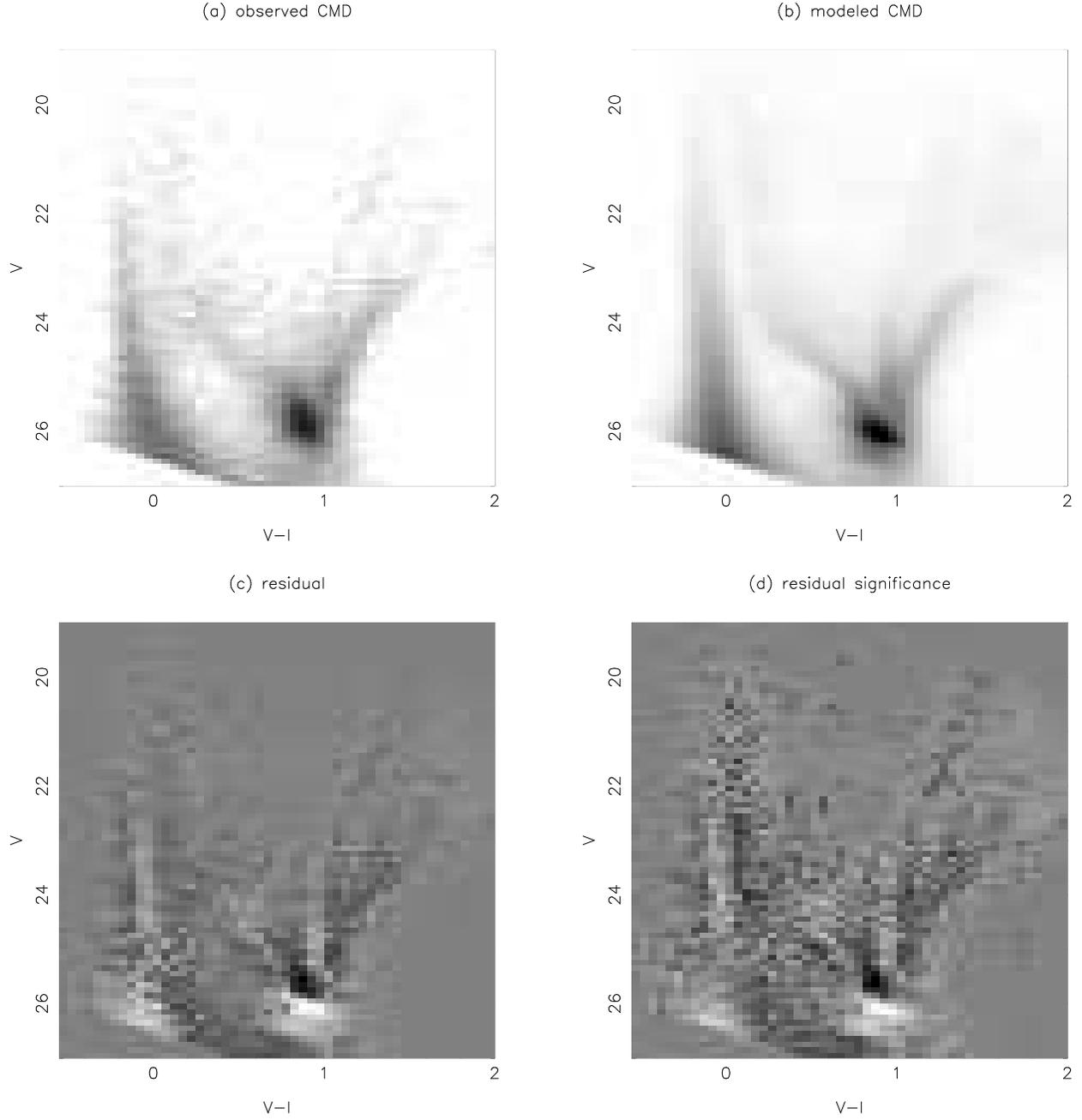}
\caption{A comparison of the observed CMD and our best-fit synthetic CMD. Panels are the same as in Figure \ref{figCMDconst}.  Note that, while still not perfect, the fit quality is vastly better. \label{figCMDbest}}
\end{figure}

The best synthetic CMD is shown in Figure \ref{figCMDbest}; the improvement over Figure \ref{figCMDconst} is readily apparent but we were still unable to perfectly fit the CMD.  This was expected, given the uncertainties in the input physics in theoretical models.  Nevertheless, we see that most parts of the CMD were fit very well; no CMD bin had an error of more than 5 $\sigma$.  Furthermore, the discrepancies are errors in position rather than errors in number of objects in a particular phase of stellar evolution.  Compared with the observed CMD, the theoretical MS is narrower, the RGB is broader, the red supergiants are bluer, and the red clump is fainter.  None of these discrepancies is sufficiently large to give us reason to believe the measured star formation history is incorrect.
\placefigure{figCMDbest}

\section{Recent Star Formation History Revisited}

Paper I presented recent (age $<0.7$ Gyr) star formation history calculations based on the MS and BHeB sequence luminosity functions.  The $V$ magnitudes along each sequence were converted into star masses, which could be transformed into star formation rates based on assumed lifetimes of those phases and an assumed initial mass function.  Summarizing those results briefly, from the BHeB stars it was determined that Sextans A had a fairly constant star formation rate (within a factor of 2) over the past 0.7 Gyr, followed by a rise (factor of 2 - 3) during the period between 50 and 100 Gyr ago, and followed by another steep rise in the last 25 Myr.  The star formation histories constructed from the MS stars were roughly consistent with this, only spanning a smaller time interval ($\sim$ 200 Myr) and showing a larger dependence on which stellar evolution models were used.  In paper I we emphasized that the method of directly converting the MS luminosity function to a recent star formation history is inherently uncertain because, unlike the BHeB stars, a position in the CMD corresponds to a large range of ages and there is significant luminosity evolution on the MS itself.  In paper I we suggested that the star formation history could be derived more accurately from the MS star using statistical methods, and thus we make that comparison here.

In our present best model we find a recent star formation history which is consistent with that found in paper I, with the notable exception that we do not see evidence of the steep increase at 25 Myr.  This feature had appeared to be quite firm as it was found in both the MS and BHeB analysis (using the Padova models, although it was not as strong and there was less agreement when the Geneva models were used).  On the other hand, the analysis of the BHeB stars is much more secure for ages greater than 25 Myr (for a given star formation rate there are more stars, and the evolution of these stars is less affected by uncertainties such as the role of stellar winds and  rotation).

To understand the cause of the discrepancy, we have made similar fits to those in paper I by adopting their distance, metallicity, and CMD sections and removing binary stars and differential extinction from our CMD models.  We find that we can reproduce the BHeB star formation history, indicating that the method used for that calculation in paper I is robust.  However, we are unable to reproduce the MS star formation history, a result we believe to stem from the treatment of evolution on the main sequence.  Since the method used in the present paper accounts for evolution, we believe the history presented here to be more accurate.

We also note that, if we model the CMD at the distance, extinction, metallicity, and binarity used in this paper but still using the restricted CMD sections (MS and BHeB stars only), we find a young star formation history consistent with that shown in Table \ref{tabSFH}, confirming the results presented here.  To clarify, our star formation history does show an elevated star formation rate during the last $\sim$ 60 Myr relative to the almost constant star formation rate from 100 - 500 Myr ago (as found in paper I), it is only that there is no increase of another factor of 2 to 4 in the last 25 Myr.  Our modeling here supports the use of the BHeB stars for reconstructing recent star formation histories, but suggests a limitation that these star formation histories be limited to parts of the CMD where the contamination of the BHeB sequence from reddened MS stars is minimal.

\section{Summary}

We have presented deep WFPC2 photometry of the dwarf irregular galaxy Sextans A, and determined its history of star formation through comparisons with synthetic CMDs.  We have made these comparisons both manually and using a numerical algorithm and find consistent results.

Sextans A began forming stars at a high rate sometime between 1 and 2.5 Gyr ago and has continuously formed stars since then.  The star formation rate over the past 0.06 Gyr is approximately three times the average over the past 2.5 Gyr, although we do not have the time resolution to determine whether similar 0.1 Gyr ``bursts'' happened more than 1 Gyr ago.  We also found a small but nonzero number of stars older than 2.5 Gyr, but the present data do not permit an accurate determination of their characteristics.  

We found very little evidence of significant chemical enrichment within Sextans A.  Based on our CMD fitting, it appears that the galaxy reached a metallicity of $\logz \approx -1.4$ more than 10 Gyr ago, and still has that metallicity today.  This metallicity is consistent with literature values from nebular spectroscopy.  An rms scatter of $\sim 0.15$ at all ages dex helps fit the observed CMD more accurately.

Our determination of the recent star formation history (age $\le$ 0.7 Gyr) agrees with that of paper I for all but the last 25 Myr.  In paper I we found a steep rise in the star formation rate that is not recovered here.  Since the present analysis demands consistency of both the MS and BHeB stars and takes into account MS luminosity evolution, we prefer the solution presented here.  This indicates that star formation  histories determined solely from BHeB stars are probably quite unreliable at magnitudes where reddened MS stars contaminate the BHeB sequence.

\acknowledgments

Support for this work was provided by NASA through grant number GO-07496 from the Space Telescope Science Institute, which is operated by AURA, Inc., under NASA contract NAS 5-26555. EDS is grateful for partial support from NASA LTSARP grant No. NAG5-9221.

\clearpage

\clearpage

\clearpage

\clearpage

\clearpage

\clearpage

\clearpage

\clearpage

\clearpage

\clearpage

\clearpage
\begin{deluxetable}{ccc}
\tablecaption{Sextans A star formation history. \label{tabSFH}}
\tablehead{
\colhead{log(Age)} &
\colhead{Age (Gyr)} &
\colhead{SFR(t) \tablenotemark{a}}}
\startdata
$ 0.0 -  7.0$ & $ 0.000 -  0.010$ & $10.7 \pm 1.7$ \\
$ 7.0 -  7.4$ & $ 0.010 -  0.025$ & $ 7.9 \pm 3.1$ \\
$ 7.4 -  7.6$ & $ 0.025 -  0.040$ & $11.8 \pm 5.3$ \\
$ 7.6 -  7.8$ & $ 0.04  -  0.06 $ & $11.7 \pm 1.5$ \\
$ 7.8 -  8.0$ & $ 0.06  -  0.10 $ & $ 3.9 \pm 1.2$ \\
$ 8.0 -  8.2$ & $ 0.10  -  0.16 $ & $ 0.7 \pm 0.5$ \\
$ 8.2 -  8.4$ & $ 0.16  -  0.25 $ & $ 1.6 \pm 0.2$ \\
$ 8.4 -  8.6$ & $ 0.25  -  0.40 $ & $ 2.4 \pm 0.3$ \\
$ 8.6 -  8.8$ & $ 0.4   -  0.6  $ & $ 1.7 \pm 0.2$ \\
$ 8.8 -  9.0$ & $ 0.6   -  1.0  $ & $ 2.7 \pm 0.3$ \\
$ 9.0 -  9.2$ & $ 1.0   -  1.6  $ & $ 0.6 \pm 0.5$ \\
$ 9.2 -  9.4$ & $ 1.6   -  2.5  $ & $ 3.5 \pm 2.2$ \\
$ 9.4 -  9.6$ & $ 2.5   -  4.0  $ & $ 0.2 \pm 0.3$ \\
$ 9.6 -  9.8$ & $ 4.    -  6.   $ & $ 0.2 \pm 0.3$ \\
$ 9.8 - 10.0$ & $ 6.    - 10.   $ & $ 0.0 \pm 0.1$ \\
$10.0 - 10.2$ & $10.    - 15.   $ & $ 1.5 \pm 1.3$ \\
\enddata
\tablenotetext{a}{Normalized to give a lifetime average star formation rate of 1.  Assuming a \citet{kro01} IMF, the lifetime average star formation rate in this field is $600 \pm 300 M_{\odot} Myr^{-1} kpc^{-2}$}
\end{deluxetable}


\begin{thebibliography}{}
\bibitem[Albert, Demers, \& Kunkel(2000)]{alb00} Albert, L., Demers, S., \& Kunkel, W. E. 2000, \aj, 119, 2780
\bibitem[Aparicio et al.(1987)]{apa87} Aparicio, A., Garc{\'\i}a-Pelayo, J. M., Moles, M., \& Melnick, J. 1987, \aaps, 71, 297
\bibitem[Aparicio \& Rodr{\'i}guez-Ulloa(1992)]{apa92} Aparicio, A. \& Rodr{\'\i}guez-Ulloa, J. A. 1992, \aap, 260, 77
\bibitem[Dohm-Palmer et al.(1997)]{doh97} Dohm-Palmer, R. C., Skillman, E. D., Saha, A., Tolstoy, E., Mateo, M., Gallagher, J. S., Hoessel, J. G., Chiosi, C. \& Defour, R. J. 1997, \aj, 114, 2514
\bibitem[Dohm-Palmer et al.(2002)]{doh02} Dohm-Palmer, R. C., Skillman, E. D., Mateo, M., Saha, A., Dolphin, A., Tolstoy, E., Gallagher, J. S., \& Cole, A. A. 2002, \aj, 123, 813 (Paper I)
\bibitem[Dolphin(2000a)]{dol00a} Dolphin, A. E. 2000a, \pasp, 112, 1383
\bibitem[Dolphin(2000b)]{dol00b} Dolphin, A. E. 2000a, \pasp, 112, 1397
\bibitem[Dolphin(2002)]{dol02} Dolphin, A. E. 2002, \mnras, 332, 91
\bibitem[Dolphin et al.(2001)]{dol01} Dolphin, A. E., Makarova, L., Karachentsev, I.D., Karachentseva, V.E., Geisler, D., Grebel, E.K., Guhathakurta, P., Hodge, P.W., Sarajedini, A., \& Seitzer, P. 2001a, \mnras, 324, 249
\bibitem[Dolphin et al.(2003)]{dol03} Dolphin, A. E., Saha, A., Skillman, E. D., Dohm-Palmer, R. C., Tolstoy, E., Cole, A. A., Gallagher, J. S., Hoessel, J. G. , \& Mateo, M. 2003, \aj in press (Paper II)
\bibitem[Girardi et al.(2000)]{gir00} Girardi, L., Bressan, A., Bertelli, G., \& Chiosi, C. 2000, \aaps, 141, 371
\bibitem[Harris, Zaritsky, \& Thompson(1997)]{har97} Harris, J., Zaritsky, D., \& Thompson, I. 1997, \aj, 114, 1933
\bibitem[Hodge, Kennicutt, \& Strobel(1994)]{hod94} Hodge, P., Kennicutt, R. C., \& Strobel, N. 1994, \pasp, 106, 765
\bibitem[Hoessel, Schommer, \& Danielson(1983)]{hoe83} Hoessel, J. G., Schommer, R. A., \& Danielson, G. E. 1983, \apj, 274, 577
\bibitem[Kroupa(2001)]{kro01} Kroupa, P. 2001, \mnras, 322, 231
\bibitem[Pettini et al.(1999)]{pet99} Pettini, M., Ellison, S. L., Steidel, C. C., \& Bowen, D. V.\ 1999, \apj, 510, 576
\bibitem[Sandage \& Carlson(1982)]{san82} Sandage, A. \& Carlson, G. 1982, \apj, 258, 439
\bibitem[Skillman, Kennicutt, \& Hodge(1989)]{ski89} Skillman, E. D., Kennicutt, R. C., \& Hodge, P. W. 1989, \apj, 347, 875
\bibitem[Skillman et al.(1988)]{ski88} Skillman, E. D., Terlevich, R., Teuben, P. J., \& van Woerden, H.\ 1988, \aap, 198, 33
\bibitem[Skillman et al.(1994)]{ski94} Skillman, E. D., Terlevich, R. J., Kennicutt, R. C., Garnett, D. R., \& Terlevich, E. 1994, \apj, 431, 172
\bibitem[Tolstoy et al.(1998)]{tol98} Tolstoy, E., Gallagher, J. S., Cole, A. A., Hoessel, J. G., Saha, A., Dohm-Palmer, R. C., Skillman, E. D., Mateom M., \& Hurley-Keller, D. 1998, \aj, 16, 1244
\bibitem[Van Dyk, Puche, \& Wong(1988)]{van88} Van Dyk, S. D., Puche, D., \& Wong, T. 1988, \aj, 116, 2341
\bibitem[Wilcots \& Hunter(2002)]{wil02} Wilcots, E. M. \& Hunter, D. A. 2002, \aj, 123, 1476
\end{thebibliography}
\end{document}